\documentstyle[12pt]{article}

\newcommand {\nn}    {\nonumber}
\newcommand {\vs}[1]  { \vspace*{#1 cm} }
\newcounter{eq}
\newcounter{sc}

\newcommand {\IJMP} {Int. J. Mod. Phys.}
\newcommand {\JHEP} {J. High Energy Phys.}

\newcommand {\NP}   {Nucl. Phys.}

\newcommand {\PL}   {Phys. Lett.}
\newcommand {\PR}   {Phys. Rev.}

\def\overleftrightarrow#1{\vbox{\ialign{##\crcr
 $\leftrightarrow$\crcr\noalign{\kern-1pt\nointerlineskip}
 $\hfil\displaystyle{#1}\hfil$\crcr}}}

\newlength{\minitwocolumn}
\begin{document}

%%%%%%%%%%%%%%%%%%%%%%%%%%%%%%%%%%%%%%%%%%%%%%%%%%%%%%%%%%%%%%%%%%
%%%%%%%%%%%%%%%%%%%%%%%% Title %%%%%%%%%%%%%%%%%%%%%%%%%%%%%%%%%%%
%%%%%%%%%%%%%%%%%%%%%%%%%%%%%%%%%%%%%%%%%%%%%%%%%%%%%%%%%%%%%%%%%%
\begin{flushright}
EDO-EP-43\\
DFPD 01/TH/34\\
hep-th/0109051\\
September, 2001\\
\end{flushright}
\vspace{20pt}

%\magnification=\magstep1
\pagestyle{empty}
\baselineskip15pt
%\font\cmssB=cmss17
%\font\cmssS=cmss10

\begin{center}
{\large\bf On the Berkovits Covariant Quantization of GS
Superstring
\vskip 1mm
}

\vspace{10mm}

Ichiro Oda
          \footnote{
          E-mail address:\ ioda@edogawa-u.ac.jp
                  }
\\

\vspace{5mm}
          Edogawa University,
          474 Komaki, Nagareyama City, Chiba 270-0198, Japan\\

\vspace{5mm}

and

\vspace{5mm}

Mario Tonin
          \footnote{
          E-mail address:\ mario.tonin@pd.infn.it
                  }
\\
\vspace{5mm}
          Dipartimento di Fisica, Universita degli Studi di Padova,\\
          Instituto Nazionale di Fisica Nucleare, Sezione di Padova,\\
          Via F. Marzolo 8, 35131 Padova, Italy\\

\end{center}

%\maketitle

\vspace{5mm}
\begin{abstract}
We study the covariant quantization of the Green-Schwarz (GS)
superstrings proposed recently by Berkovits.
In particular, we reformulate the Berkovits approach in a way 
that clarifies its relation with the GS approach and allows to derive 
in a straightforward way its extension to curved spacetime background.
We explain the procedure working explicitly in the case of the heterotic 
string.

\end{abstract}

\newpage
\pagestyle{plain}
\pagenumbering{arabic}
%\setcounter{page}{1}

%%%%%%%%%%%%%%%%%%%%%%%%%%%%%%%%%%%%%%%%%%%%%%%%%%%%%%%%%%%%%%%%%%
%%%%%%%%%%%%%%%%%%%%%%%% Article %%%%%%%%%%%%%%%%%%%%%%%%%%%%%%%%%
%%%%%%%%%%%%%%%%%%%%%%%%%%%%%%%%%%%%%%%%%%%%%%%%%%%%%%%%%%%%%%%%%%

\rm
%%%%%%%%%%%%%%%%%%%%%%%%%%%%%%%%%%%%%%%%%%%%%%%%%%%%%%%%%%%%%%%%%%%%%
%%%%%%%%%%%%%%%%%%%%%%%%%%%%%%   SEC  1    %%%%%%%%%%%%%%%%%%%%%%%%%%
%%%%%%%%%%%%%%%%%%%%%%%%%%%%%%%%%%%%%%%%%%%%%%%%%%%%%%%%%%%%%%%%%%%%%
\section{Introduction}

  Notably with the advent of the Green-Schwarz (GS) superstring action
with a manifest space-time supersymmetry \cite{GS}, there have been
a lot of efforts to quantize the action in a Lorentz covariant manner.
However no one has succeeded in making a fully  covariant quantization 
of the GS superstring  action. The source of the difficulty is well
known, that is, it is impossible to achieve the desired separation
of fermionic first class and second class constraints associated with
local $\kappa$ symmetry in a manifestly covariant way.
As in ten dimensions the smallest covariant spinor corresponding to 
a Majorana-Weyl spinor has 16 real components, 8 first class and 
8 second class constraints that arise in heterotic or type I  GS 
superstrings do not fit into such covariant spinor representation 
separately.
For Type II GS superstrings the same
happens in each of the two, left-handed or right-handed, sectors.

  If one tries to perform the quantization following the standard BRST-BV
recipe, one ends with an infinite set of ghosts and ghosts of ghosts, that is,
$\kappa$ symmetry is infinitely reducible.
All attempts \cite{a,b,c},
to extract from this situation a consistent quantization scheme failed,
leading to a BRS charge with the wrong cohomology.
A way to perform at the classical level a covariant separation
of first and second class constraints is the Lorentz  harmonic
approach \cite{Bandos}, but nobody has succeded in getting a
workable quantization scheme along this line.

  Recently, Berkovits has proposed an interesting approach to covariant
quantization of superstrings,  using pure
spinors \cite{Ber1,Ber2, Ber3, Ber4}. The starting point of this approach is
the BRS charge $Q_{BRS} = \oint \lambda^\alpha d_\alpha$ where
$\lambda^\alpha$ are pure spinors satisfying the equation $\lambda^\alpha
\Gamma^m_{\alpha\beta} \lambda^\beta = 0$ and $d_\alpha \approx 0$
denote the GS fermionic constraints. The action is the free field action
involving the superspace coordinates $X^m$ and $\theta^\mu$, the conjugate
momenta of the Grassmann coordinate $\theta$, the pure spinor ghost $\lambda$
and its conjugate momentum.
In this approach the central charge vanishes, the
BRS charge is nilpotent and has the same cohomology as
the BRS charge of the Neveu-Schwarz-Ramond (NSR) formalism \cite{Neveu}.
Moreover  vertices can be constructed which, modulo a very plausible
conjecture, give the correct tree amplitudes.

  The Berkovits approach appears to be in the right direction
for covariant quantization of the Green-Schwarz superstring action,
but the method used there is not conventional. For instance, the BRS charge
$Q_{BRS} = \oint \lambda^\alpha d_\alpha$ contains both first class
and second class constraints, whereas the conventional BRS charge
involves only first class constraints. One of the motivations of this paper
is to fill the gap between the Berkovits approach and the conventional
BRS approach in order to clarify the relation between this approach and the
GS one. To be definite we shall consider only the case of the heterotic
string.
The other cases can be treated similarly.

  This article is organized as follows.
In section two, we review briefly the Green-Schwarz superstring action,
pure spinors, the $SO(1,9)/U(5)$ coset formalism and the Berkovits
approach.
In section three, in a flat background we introduce a modification of
the GS action to get a BRS-invariant action,
from which the Berkovits action is derived by a standard BRS procedure.
Moreover,
in section four, the formulation used in section three is generalized to
the case of curved background. The final section is devoted to discussions.

%%%%%%%%%%%%%%%%%%%%%%%%%%%%%%%%%%%%%%%%%%%%%%%%%%%%%%%%%%%%%%%%%%%%%
%%%%%%%%%%%%%%%%%%%%%%%%%%%%%%   SEC  2    %%%%%%%%%%%%%%%%%%%%%%%%%%
%%%%%%%%%%%%%%%%%%%%%%%%%%%%%%%%%%%%%%%%%%%%%%%%%%%%%%%%%%%%%%%%%%%%%
\section{Review}

Before presenting our results, we shall review the salient points of
the superspace formulation of the Green-Schwarz heterotic
superstring action, pure spinors, the
$SO(1,9)/U(5)$ coset formalism and the Berkovits action, which will be
fully utilized in later sections.

We start with the superspace formulation of the Green-Schwarz
heterotic superstring action in a general curved space-time:
%**   2.1 %%%%%%%%%%%%%%%%%%%%%%%%%%%%%%%%%%%%%%%%%%%%%%%%%%%%%%%%%
\begin{eqnarray}
I_{GS} = \frac{1}{2} \int_{M_2} \det e \ e^\varphi \
E^a_+ E_{-a} + \int_{M_2} B_2 + \sum_I \int_{M_2} \psi^I
{\cal{D}}_- \psi^I,
\label{2.1}
\end{eqnarray}
%%%%%%%%%%%%%%%%%%%%%%%%%%%%%%%%%%%%%%%%%%%%%%%%%%%%%%%%%%%%%%%%%%%
where $M_2$ denotes the two-dimensional world sheet, $e^{\pm}_i$
(with its inverse $e^i_{\pm}$) are world sheet vielbeins, $E^a_{\pm}$
are the pullback of the superspace vielbeins, $B_2$ is the NS-NS two
form potential and $\varphi$ is the dilaton. Concretely,
the pullback of the supervielbeins $E^A_{\pm}$
can be expressed in terms of the superspace variables $Z^M =
(X^m, \theta^\mu)$ by $E^A_{\pm} = e^i_{\pm} \partial_i Z^M E^A_M(Z)$.
The Latin letters are used for vectors, while the Greek ones
are for spinors and the Capital letters for both. Moreover, the letters
from the beginning of the
alphabet are tangent space indices, whereas the letters from the middle
are target space indices. Finally, the last term in the right hand side
in Eq. (\ref{2.1}) denotes a set of left-moving heterotic fermions
where the covariant derivative is defined as  ${\cal{D}}_- = \partial_-
+ \partial_- Z^M A_M$ with $A = dZ^M A_M$ being the one-form gauge
potentials.

It is well known that the Green-Schwarz action (\ref{2.1}) is invariant
under local $\kappa$ symmetry \cite{Siegel1} only when the background
satisfies the SUGRA-SYM background constraints \cite{Grisaru}.
Indeed, under the local $\kappa$ symmetry
%**   2.2 %%%%%%%%%%%%%%%%%%%%%%%%%%%%%%%%%%%%%%%%%%%%%%%%%%%%%%%%%
\begin{eqnarray}
\delta Z^M E^\alpha_M = w^\alpha = E^a_- \Gamma_a^{\alpha\beta}
\kappa_\beta, \ \delta Z^M E^a_M = 0,
\label{2.2}
\end{eqnarray}
%%%%%%%%%%%%%%%%%%%%%%%%%%%%%%%%%%%%%%%%%%%%%%%%%%%%%%%%%%%%%%%%%%%
the action transforms as
%**   2.3 %%%%%%%%%%%%%%%%%%%%%%%%%%%%%%%%%%%%%%%%%%%%%%%%%%%%%%%%%
\begin{eqnarray}
\delta I_{GS} = - \int_{M_2} \det e \ w E^a_- \Gamma_a \hat{E}_+,
\label{2.3}
\end{eqnarray}
%%%%%%%%%%%%%%%%%%%%%%%%%%%%%%%%%%%%%%%%%%%%%%%%%%%%%%%%%%%%%%%%%%%
where the SUGRA-SYM background constraints have been used and
we have defined
%**   2.4 %%%%%%%%%%%%%%%%%%%%%%%%%%%%%%%%%%%%%%%%%%%%%%%%%%%%%%%%%
\begin{eqnarray}
\hat{E}^\alpha_+ = ( E^\alpha_+ - \frac{1}{2} E^a_+ \Gamma_a^{\alpha\beta}
D_\beta \varphi) e^\varphi.
\label{2.4}
\end{eqnarray}
%%%%%%%%%%%%%%%%%%%%%%%%%%%%%%%%%%%%%%%%%%%%%%%%%%%%%%%%%%%%%%%%%%%
Then, provided that the symmetry (\ref{2.2}) is supplemented with
$\delta e^i_+ = 2 \kappa \hat{E}_+ e^i_-$ and $\delta e^i_- = 0$,
the Green-Schwarz action becomes invariant, $\delta I_{GS} = 0$
under the local $\kappa$ symmetry.

We now turn our attention to the case of a flat background in conformal gauge.
Then, the heterotic action (\ref{2.1}) reduces to the form
%**   2.5 %%%%%%%%%%%%%%%%%%%%%%%%%%%%%%%%%%%%%%%%%%%%%%%%%%%%%%%%%
\begin{eqnarray}
I_{GS} = \int d^2 z \ \Big[ \frac{1}{2} \Pi^m \bar{\Pi}_m
+ \frac{1}{4} ( \Pi^m \theta \Gamma_m \bar{\partial} \theta
- \bar{\Pi}^m \theta \Gamma_m \partial \theta) \Big] +
\sum_I \int \psi^I \partial \psi^I.
\label{2.5}
\end{eqnarray}
%%%%%%%%%%%%%%%%%%%%%%%%%%%%%%%%%%%%%%%%%%%%%%%%%%%%%%%%%%%%%%%%%%%
In this case, $E^A_{\pm} = (E^a_{\pm}, E^\alpha_{\pm})$ are of form
%**   2.6 %%%%%%%%%%%%%%%%%%%%%%%%%%%%%%%%%%%%%%%%%%%%%%%%%%%%%%%%%
\begin{eqnarray}
&{}& E^a_- \rightarrow \Pi^m = \partial X^m + \frac{1}{2} \theta
\Gamma^m \partial \theta, \nn\\
&{}& E^a_+ \rightarrow \bar{\Pi}^m = \bar{\partial} X^m + \frac{1}{2}
\theta \Gamma^m \bar{\partial} \theta, \nn\\
&{}& E^\alpha_{\pm} = \partial_{\pm} \theta^\alpha.
\label{2.6}
\end{eqnarray}
%%%%%%%%%%%%%%%%%%%%%%%%%%%%%%%%%%%%%%%%%%%%%%%%%%%%%%%%%%%%%%%%%%%

% Note that we employ a bimodular representation of the ten-dimensional
%$\gamma$ matrices where $32 \times 32 \ \gamma$ matrices are divided into
%two $16 \times 16 \ \Gamma$ matrices which are symmetric with respect
%to exchange of spinor indices, $\Gamma^m_{\alpha\beta} =
%\Gamma^m_{\beta\alpha}$.

As usual $\Gamma^m$ are the Dirac matrices $\gamma^m$ times the charge
conjugation matrix and are $16\times 16$ matrices symmetric with respect
to exchange of spinor indices, $\Gamma^m_{\alpha\beta} =
\Gamma^m_{\beta\alpha}$.
Moreover we shall use the notation $\Gamma^{m_1 \cdots m_p}$
to denote the antisymmetric product of $p$ $\gamma$ times the charge 
conjugation.

This action (\ref{2.5}) possesses the Virasoro constraint $\Pi^m
\Pi_m \approx 0$ and fermionic constraints $d_\alpha \equiv
p_\alpha - \frac{1}{2} (\Pi^m - \frac{1}{4} \theta \Gamma^m \partial \theta)
(\Gamma_m \theta)_\alpha \approx 0$ where $p_\alpha$ are the canonical
momenta conjugate to $\theta^\alpha$. The latter constraints include 8 first
class constraints and 8 second class ones, a fact which is the
source of the difficulty of covariant quantization as mentioned above.
In what follows, the left-moving heterotic fermions play no role
and therefore will be ignored for simplicity.

It is worthwhile to point out that there is an interesting identity
by Siegel \cite{Siegel2}, which is given by
%**   2.7 %%%%%%%%%%%%%%%%%%%%%%%%%%%%%%%%%%%%%%%%%%%%%%%%%%%%%%%%%
\begin{eqnarray}
\int d^2 z \ \Big[ \frac{1}{2} \partial X^m \bar{\partial} X_m
+ p_\alpha \bar{\partial} \theta^\alpha \Big]
= I_{GS} + \int d^2 z \ d_\alpha \bar{\partial} \theta^\alpha.
\label{2.7}
\end{eqnarray}
%%%%%%%%%%%%%%%%%%%%%%%%%%%%%%%%%%%%%%%%%%%%%%%%%%%%%%%%%%%%%%%%%%%
With the OPE's
%**   2.8 %%%%%%%%%%%%%%%%%%%%%%%%%%%%%%%%%%%%%%%%%%%%%%%%%%%%%%%%%
\begin{eqnarray}
X^m(y) X^n(z) \rightarrow - \eta^{mn} \log |y-z|^2, \
p_\alpha(y) \theta^\beta(z) \rightarrow \frac{1}{y-z}
\delta_\alpha^\beta,
\label{2.8}
\end{eqnarray}
%%%%%%%%%%%%%%%%%%%%%%%%%%%%%%%%%%%%%%%%%%%%%%%%%%%%%%%%%%%%%%%%%%%
one can calculate the OPE among the fermionic constraints $d_\alpha
\approx 0$
%**   2.9 %%%%%%%%%%%%%%%%%%%%%%%%%%%%%%%%%%%%%%%%%%%%%%%%%%%%%%%%%
\begin{eqnarray}
d_\alpha(y) d_\beta(z) \rightarrow - \frac{1}{y-z} \Pi^m
(\Gamma_m)_{\alpha\beta}.
\label{2.9}
\end{eqnarray}
%%%%%%%%%%%%%%%%%%%%%%%%%%%%%%%%%%%%%%%%%%%%%%%%%%%%%%%%%%%%%%%%%%%

Here let us introduce the concept of the "pure spinors" which plays an
important role in the Berkovits works \cite{Ber1,Ber2, Ber3, Ber4}.
(See also related works \cite{Tonin2, Ber5}.)
Pure spinors are simply defined as complex, commuting, Weyl spinors
such that
%**   2.10 %%%%%%%%%%%%%%%%%%%%%%%%%%%%%%%%%%%%%%%%%%%%%%%%%%%%%%%%%
\begin{eqnarray}
\lambda^\alpha
\Gamma^m_{\alpha\beta} \lambda^\beta = 0.
\label{2.10}
\end{eqnarray}
%%%%%%%%%%%%%%%%%%%%%%%%%%%%%%%%%%%%%%%%%%%%%%%%%%%%%%%%%%%%%%%%%%%
{}From this definition and Eq. (\ref{2.9}), it turns out that
the BRS charge
%**   2.11 %%%%%%%%%%%%%%%%%%%%%%%%%%%%%%%%%%%%%%%%%%%%%%%%%%%%%%%%%
\begin{eqnarray}
Q_{BRS} = \oint \lambda^\alpha d_\alpha,
\label{2.11}
\end{eqnarray}
%%%%%%%%%%%%%%%%%%%%%%%%%%%%%%%%%%%%%%%%%%%%%%%%%%%%%%%%%%%%%%%%%%%
becomes nilpotent $Q^2_{BRS} = 0$. At this stage, we wish to
mention one important remark.
The hermiticity condition on the BRS charge automatically leads to
the hermiticity condition on the pure spinors $\lambda^\alpha$
%**   2.12 %%%%%%%%%%%%%%%%%%%%%%%%%%%%%%%%%%%%%%%%%%%%%%%%%%%%%%%%%
\begin{eqnarray}
\lambda^\dagger = \lambda,
\label{2.12}
\end{eqnarray}
%%%%%%%%%%%%%%%%%%%%%%%%%%%%%%%%%%%%%%%%%%%%%%%%%%%%%%%%%%%%%%%%%%%
which must be imposed at the quantum level. On the other hand,
as classical fields, the pure spinors $\lambda$ are complex, and
using $\Gamma^0 = 1$ the time component of Eq. (\ref{2.10}) gives
%**   2.13 %%%%%%%%%%%%%%%%%%%%%%%%%%%%%%%%%%%%%%%%%%%%%%%%%%%%%%%%%
\begin{eqnarray}
\lambda^2 = 0.
\label{2.13}
\end{eqnarray}
%%%%%%%%%%%%%%%%%%%%%%%%%%%%%%%%%%%%%%%%%%%%%%%%%%%%%%%%%%%%%%%%%%%
Then, Eqs.  (\ref{2.12}) and  (\ref{2.13}) are not inconsistent at the
quantum level since the pure spinors $\lambda$ reside in a Hilbert
space with indefinite metric.

As a final preparation for our purpose, let us explain the coset
$SO(1,9)/U(5)$. $U(5)$ is a subgroup of $SO(1,9)$ which acts linearly
on $X^r = X^{2r-2} + i X^{2r-1}$ (as well as $\bar{X}^r = X^{2r-2} -
i X^{2r-1}$) as $X'^r = \Lambda^r_s X^s$ where $\Lambda \in U(5)$ and
$r, s = 1, 2, \cdots, 5$. A spinor can be expressed in a basis of
eigenvectors of the 5 commuting $SO(1,9)$ generators $\frac{1}{2 i}
\Gamma^{2r-2} \Gamma^{2r-1}$ as $\phi^\alpha \equiv |\pm \pm \pm \pm \pm>$.
 Then, complex Weyl spinors have an even number of "$-$"
eigenvalues and are decomposed into irreducible representations of
$U(5)$ as
%**   2.14 %%%%%%%%%%%%%%%%%%%%%%%%%%%%%%%%%%%%%%%%%%%%%%%%%%%%%%%%%
\begin{eqnarray}
|+ + + + +>                 &\rightarrow& \phi^0, \nn\\
|+ + - - +> + \ permutations  &\rightarrow& \phi_{[rs]}, \nn\\
|+ - - - -> + \ permutations  &\rightarrow& \phi^r,
\label{2.14}
\end{eqnarray}
%%%%%%%%%%%%%%%%%%%%%%%%%%%%%%%%%%%%%%%%%%%%%%%%%%%%%%%%%%%%%%%%%%%
where each representation transforms respectively as $(\bf{1},
\bar{\bf{10}}, \bf{5})$.
{}For pure spinors $\lambda^\alpha$  we have the relation \cite{Ber1}
%**   2.15 %%%%%%%%%%%%%%%%%%%%%%%%%%%%%%%%%%%%%%%%%%%%%%%%%%%%%%%%%
\begin{eqnarray}
\lambda^\alpha = \Big( \lambda^0, \ \lambda_{[rs]}, \ \lambda^r
= - \frac{1}{8 \lambda^0} \varepsilon^{r s_1 s_2 s_3 s_4}
\lambda_{[s_1 s_2]} \lambda_{[s_3 s_4]} \Big).
\label{2.15}
\end{eqnarray}
%%%%%%%%%%%%%%%%%%%%%%%%%%%%%%%%%%%%%%%%%%%%%%%%%%%%%%%%%%%%%%%%%%%
and therefore a pure spinor has eleven degrees of freedom.

 It is convenient to define the constant "harmonics" 
  $(v^0_\alpha, v_{[rs]\alpha}, v^r_\alpha)$ that take out the 
  $U(5)$ representations of an  $SO(1,9)$ Weyl spinor, that is:
%**   2.16 %%%%%%%%%%%%%%%%%%%%%%%%%%%%%%%%%%%%%%%%%%%%%%%%%%%%%%%%%
\begin{eqnarray}
\phi^0 = v^0_\alpha \phi^\alpha, \ \phi_{[rs]} = v_{[rs]\alpha}
\phi^\alpha, \ \phi^r = v^r_\alpha \phi^\alpha.
\label{2.16}
\end{eqnarray}
%%%%%%%%%%%%%%%%%%%%%%%%%%%%%%%%%%%%%%%%%%%%%%%%%%%%%%%%%%%%%%%%%%%
Of course, in a similar way, we can describe the anti-Weyl spinor
by means of $(\bar{v}_0^\alpha, \bar{v}^{[rs]\alpha}, \bar{v}_r^\alpha)$.
Here let us introduce $\omega_\alpha$ which are the "almost" conjugate
momenta of $\lambda$ with the OPE:
%**   2.17 %%%%%%%%%%%%%%%%%%%%%%%%%%%%%%%%%%%%%%%%%%%%%%%%%%%%%%%%%
\begin{eqnarray}
\omega_\alpha(y) \lambda^\beta(z) &\rightarrow& - \frac{1}{y-z}
\Big[ \delta_\alpha^\beta - \frac{1}{2} \frac{(\Gamma^m \lambda)_\alpha
(v^0 \Gamma_m)^\beta}{v^0 \lambda}  \Big], \nn\\
&\equiv& - \frac{1}{y-z} [ \delta_\alpha^\beta - K_\alpha{}^\beta],
\label{2.17}
\end{eqnarray}
%%%%%%%%%%%%%%%%%%%%%%%%%%%%%%%%%%%%%%%%%%%%%%%%%%%%%%%%%%%%%%%%%%%
where $K_\alpha{}^\beta$ and $\delta_\alpha^\beta - K_\alpha{}^\beta$
are projectors and $\lambda^\alpha K_\alpha{}^\beta = 0$. The projector
$K_\alpha{}^\beta$ in the OPE (\ref{2.17}) is needed in order that 
$\omega_\alpha$ should be consistent with the pure spinor
condition (\ref{2.10}), i.e.,
$\omega_\alpha(y) \lambda \Gamma^m \lambda(z) \rightarrow 0$.
Moreover, it is useful to introduce the tensor operators
%**   2.18 %%%%%%%%%%%%%%%%%%%%%%%%%%%%%%%%%%%%%%%%%%%%%%%%%%%%%%%%%
\begin{eqnarray}
N^{mn} = \frac{1}{2} \omega \Gamma^{mn} \lambda, \
N_\alpha{}^\beta = N^{mn} \frac{1}{4} (\Gamma_{mn})_\alpha{}^\beta,
\label{2.18}
\end{eqnarray}
%%%%%%%%%%%%%%%%%%%%%%%%%%%%%%%%%%%%%%%%%%%%%%%%%%%%%%%%%%%%%%%%%%%
which satisfy the OPE of the $SO(1,9)$ Lorentz generator densities
up to a central charge. The total Lorentz generator densities
 $M^{mn} = L^{mn} + N^{mn}$
have the same central charge as in NSR formalism.

With these facts in mind, Berkovits has considered the action
in a flat background
%**   2.19 %%%%%%%%%%%%%%%%%%%%%%%%%%%%%%%%%%%%%%%%%%%%%%%%%%%%%%%%%
\begin{eqnarray}
I_B =  \int d^2 z \ \Big[ \frac{1}{2} \partial X^m \bar{\partial} X_m
+ p_\alpha \bar{\partial} \theta^\alpha + \omega_0 \bar{\partial}
\lambda^0 + \frac{1}{2} \omega^{[rs]} \bar{\partial}
\lambda_{[rs]} \Big],
\label{2.19}
\end{eqnarray}
%%%%%%%%%%%%%%%%%%%%%%%%%%%%%%%%%%%%%%%%%%%%%%%%%%%%%%%%%%%%%%%%%%%
and shown that the total central charge vanishes, $Q_{BRS}$ has
the same cohomology as the BRS charge of NSR formalism, and vertex
operators yield the correct tree amplitudes \cite{Ber1,Ber2, Ber3, Ber4}.

%%%%%%%%%%%%%%%%%%%%%%%%%%%%%%%%%%%%%%%%%%%%%%%%%%%%%%%%%%%%%%%%%%%%%
%%%%%%%%%%%%%%%%%%%%%%%%%%%%%%   SEC  3    %%%%%%%%%%%%%%%%%%%%%%%%%%
%%%%%%%%%%%%%%%%%%%%%%%%%%%%%%%%%%%%%%%%%%%%%%%%%%%%%%%%%%%%%%%%%%%%%
\section{New presentation of the Berkovits approach in flat background}

In previous section we have discussed the Berkovits works briefly.
Even if his formalism has many good properties as  mentioned at
the end of the section, it has some unusual features. In particular, the
BRS charge $Q_{BRS}$, (\ref{2.11}) is composed of the constraints
$d_\alpha \approx 0$, which contain not only first class but also
second class constraints, whereas the conventional BRS charge is entirely
composed of first class constraints. In addition and related to it, his
action (\ref{2.19}) cannot be obtained from the Green-Schwarz action by
the '$\it{standard}$' BRS procedure. Here by '$\it{standard}$' BRS procedure
we mean  that one starts with an invariant action and then adds to the
action the gauge fixing term plus the FP ghost term which are written together
as $\{Q_{BRS}, \Psi\}$ where $\Psi$ is the so-called '$\it{gauge \ fermion}$'
with ghost number $-1$.
In this section, we shall construct a BRS-invariant action starting from
the GS one  and derive the Berkovits action by adding to it the BRS
transformation of a gauge fermion. We shall limit ourselves to
the Green-Schwarz heterotic superstring action in a flat
background. The case of a general curved background will be
treated in next section.

In fact, the Green-Schwarz action $I_{GS}$ in Eq. (\ref{2.5}) in
a flat background space-time is not invariant under the BRS
transformation generated by $Q_{BRS}$, (\ref{2.11}) and
the variation takes the form
%**   3.1 %%%%%%%%%%%%%%%%%%%%%%%%%%%%%%%%%%%%%%%%%%%%%%%%%%%%%%%%%
\begin{eqnarray}
\delta I_{GS} =  \int d^2 z \ \lambda \Gamma^m \Pi_m \bar{\partial} \theta,
\label{3.1}
\end{eqnarray}
%%%%%%%%%%%%%%%%%%%%%%%%%%%%%%%%%%%%%%%%%%%%%%%%%%%%%%%%%%%%%%%%%%%
where we have used the OPE's in Eq. (\ref{2.8}). Note here that
this result (\ref{3.1}) precisely corresponds to Eq. (\ref{2.3})
(an additional
$-1$ factor does not appear in (\ref{3.1}) compared to (\ref{2.3})
owing to the bosonic character of pure spinors $\lambda$).
% (In (\ref{2.3}),
%the gauge parameter $w$ is a fermionic one, so $-1$ factor appears
%in moving $w$ to the leftmost side.)

The key idea is to add to $I_{GS}$ a new term $I_{new}$ so that
%**   3.2 %%%%%%%%%%%%%%%%%%%%%%%%%%%%%%%%%%%%%%%%%%%%%%%%%%%%%%%%%
\begin{eqnarray}
I_0 \equiv I_{GS} + I_{new},
\label{3.2}
\end{eqnarray}
%%%%%%%%%%%%%%%%%%%%%%%%%%%%%%%%%%%%%%%%%%%%%%%%%%%%%%%%%%%%%%%%%%%
is invariant under the BRS transformation. Is it possible to find such a
new term ? We can see that the following expression works well.
Actually, provided that we take
%**   3.3 %%%%%%%%%%%%%%%%%%%%%%%%%%%%%%%%%%%%%%%%%%%%%%%%%%%%%%%%%
\begin{eqnarray}
I_{new} = - \frac{1}{2} \int d^2 z \ \frac{(\bar{\partial} \theta
\Gamma^m \lambda) (v^0 \Gamma_m d)}{v^0 \lambda},
\label{3.3}
\end{eqnarray}
%%%%%%%%%%%%%%%%%%%%%%%%%%%%%%%%%%%%%%%%%%%%%%%%%%%%%%%%%%%%%%%%%%%
by means of Eqs. (\ref{2.9}), (\ref{2.10}) and the Fierz identity
$\Gamma^m{}_{\alpha (\beta} \Gamma^m{}_{\rho \sigma)} = 0$, we find
%**   3.4 %%%%%%%%%%%%%%%%%%%%%%%%%%%%%%%%%%%%%%%%%%%%%%%%%%%%%%%%%
\begin{eqnarray}
\delta I_{new} = - \int d^2 z \ \lambda \Gamma^m \Pi_m \bar{\partial}
\theta.
\label{3.4}
\end{eqnarray}
%%%%%%%%%%%%%%%%%%%%%%%%%%%%%%%%%%%%%%%%%%%%%%%%%%%%%%%%%%%%%%%%%%%
As a result, the action $I_0$ is BRS-invariant, $\delta I_0 = 0$.

Since we have constructed a BRS-invariant action, we are now ready
to apply the '$\it{standard}$' BRS recipe. The appropriate choice of
gauge fermion is given by
%**   3.5 %%%%%%%%%%%%%%%%%%%%%%%%%%%%%%%%%%%%%%%%%%%%%%%%%%%%%%%%%
\begin{eqnarray}
\Psi = \int d^2 z \ \omega_\alpha \bar{\partial} \theta^\alpha.
\label{3.5}
\end{eqnarray}
%%%%%%%%%%%%%%%%%%%%%%%%%%%%%%%%%%%%%%%%%%%%%%%%%%%%%%%%%%%%%%%%%%%
Then, adding this BRS variation to the BRS-invariant action $I_0$,
we obtain a "gauge-fixed", BRS-invariant action
%**   3.6 %%%%%%%%%%%%%%%%%%%%%%%%%%%%%%%%%%%%%%%%%%%%%%%%%%%%%%%%%
\begin{eqnarray}
I &=& I_0 + \delta \Psi, \nn\\
&=& \int d^2 z \ \Big[ \frac{1}{2} \partial X^m \bar{\partial} X_m
+ p_\alpha \bar{\partial} \theta^\alpha + \omega_\alpha \bar{\partial}
\lambda^\alpha \Big].
\label{3.6}
\end{eqnarray}
%%%%%%%%%%%%%%%%%%%%%%%%%%%%%%%%%%%%%%%%%%%%%%%%%%%%%%%%%%%%%%%%%%%
Here the last term in the integrand can be rewritten as
%**   3.7 %%%%%%%%%%%%%%%%%%%%%%%%%%%%%%%%%%%%%%%%%%%%%%%%%%%%%%%%%
\begin{eqnarray}
\omega_\alpha \bar{\partial} \lambda^\alpha &=&
\omega_0 \bar{\partial} \lambda^0 + \frac{1}{2} \omega^{[rs]} \bar{\partial}
\lambda_{[rs]} + \omega_r \bar{\partial}
\Big( - \frac{1}{8 \lambda^0} \varepsilon^{r s_1 s_2 s_3 s_4}
\lambda_{[s_1 s_2]} \lambda_{[s_3 s_4]} \Big), \nn\\
&=& \omega'_0 \bar{\partial} \lambda^0 + \frac{1}{2} \omega'^{[rs]} 
\bar{\partial} \lambda_{[rs]},
\label{3.7}
\end{eqnarray}
%%%%%%%%%%%%%%%%%%%%%%%%%%%%%%%%%%%%%%%%%%%%%%%%%%%%%%%%%%%%%%%%%%%
where
%**   3.8 %%%%%%%%%%%%%%%%%%%%%%%%%%%%%%%%%%%%%%%%%%%%%%%%%%%%%%%%%
\begin{eqnarray}
\omega'_0 &=& \omega_0 + \frac{1}{8 (\lambda^0)^2}
\varepsilon^{r s_1 s_2 s_3 s_4}
\omega_r \lambda_{[s_1 s_2]} \lambda_{[s_3 s_4]}, \nn\\
\omega'^{[rs]} &=& \omega^{[rs]} - \frac{1}{4 \lambda^0}
\varepsilon^{t t_1 t_2 r s} \omega_t \lambda_{[t_1 t_2]}.
\label{3.8}
\end{eqnarray}
%%%%%%%%%%%%%%%%%%%%%%%%%%%%%%%%%%%%%%%%%%%%%%%%%%%%%%%%%%%%%%%%%%%
Thus, modulo the field redefinitions of $\omega$, which is harmless,
the "gauge-fixed", BRS-invariant action $I$ precisely coincides with
the Berkovits action (\ref{2.19}).

%%%%%%%%%%%%%%%%%%%%%%%%%%%%%%%%%%%%%%%%%%%%%%%%%%%%%%%%%%%%%%%%%%%%%
%%%%%%%%%%%%%%%%%%%%%%%%%%%%%%   SEC  4    %%%%%%%%%%%%%%%%%%%%%%%%%%
%%%%%%%%%%%%%%%%%%%%%%%%%%%%%%%%%%%%%%%%%%%%%%%%%%%%%%%%%%%%%%%%%%%%%
\section{Generalization to curved background}

In previous section, we have considered only the case of a flat
background space-time. Now we move on to the construction of
the Berkovits action in a curved background.
Our presentation of the Berkovits approach allows to derive it in a quite 
straightforward and clean way.

As mentioned in section two, the Green-Schwarz action is invariant under
local $\kappa$ symmetry only when the background satisfies the SUGRA-SYM
background constraints \cite{Grisaru}. A standard set of constraints
is given by \cite{Nilsson, Witten}
%**   4.1 %%%%%%%%%%%%%%%%%%%%%%%%%%%%%%%%%%%%%%%%%%%%%%%%%%%%%%%%%
\begin{eqnarray}
&{}& T^a_{\alpha\beta} -  \Gamma^a_{\alpha\beta} = T^a_{\alpha b}
= T^\alpha_{\beta\gamma} = 0, \nn\\
&{}& H_{\alpha\beta\gamma} = 0 = H_{\alpha\beta a} - \frac{1}{2} e^\varphi
(\Gamma_a)_{\alpha\beta}, \nn\\
&{}& F_{\alpha\beta} = 0,
\label{4.1}
\end{eqnarray}
%%%%%%%%%%%%%%%%%%%%%%%%%%%%%%%%%%%%%%%%%%%%%%%%%%%%%%%%%%%%%%%%%%%
where $T^A = D E^A$ is the superspace torsion, and $H = d B$ and
$F = d A + A^2$ are respectively the curvatures of $B$ field and
gauge fields. Note that at this level, SYM is completely decoupled
from the $B$ field, and the Chaplin-Manton coupling arises from
$\sigma$-model loop corrections in order to cancel anomalies
associated with the $\kappa$ symmetry in the Green-Schwarz
formulation \cite{Atick1, Tonin3}.
The constraints (\ref{4.1}) then lead to \cite{Atick2}
%**   4.2 %%%%%%%%%%%%%%%%%%%%%%%%%%%%%%%%%%%%%%%%%%%%%%%%%%%%%%%%%
\begin{eqnarray}
T^\alpha_{a \beta} = - \frac{1}{24} ( \Gamma_a \Gamma^{f_1 f_2 f_3}
)_\beta{}^\alpha T_{f_1 f_2 f_3}, \ H_{ab \alpha} = - \frac{1}{2}
e^\varphi ( \Gamma_{a b} )_\alpha{}^\beta D_\beta \varphi,
\label{4.2}
\end{eqnarray}
%%%%%%%%%%%%%%%%%%%%%%%%%%%%%%%%%%%%%%%%%%%%%%%%%%%%%%%%%%%%%%%%%%%
where
%**   4.3 %%%%%%%%%%%%%%%%%%%%%%%%%%%%%%%%%%%%%%%%%%%%%%%%%%%%%%%%%
\begin{eqnarray}
D_\alpha D_\beta \varphi + D_\alpha \varphi D_\beta \varphi
+\frac{1}{2} \Gamma^a_{\alpha\beta} D_a \varphi = - \frac{1}{12} 
(\Gamma^{f_1 f_2 f_3})_{\alpha\beta} T_{f_1 f_2 f_3}.
\label{4.3}
\end{eqnarray}
%%%%%%%%%%%%%%%%%%%%%%%%%%%%%%%%%%%%%%%%%%%%%%%%%%%%%%%%%%%%%%%%%%%

Now the Green-Schwarz action is given by  (\ref{2.1}) taken in 
conformal gauge and
%the constraints are the Virasoro and
the fermionic constraints are 
%**   4.4 %%%%%%%%%%%%%%%%%%%%%%%%%%%%%%%%%%%%%%%%%%%%%%%%%%%%%%%%%
\begin{eqnarray}
d_\alpha \equiv p_\alpha  - \frac{1}{2} ( E^a_- B_{a \alpha} +
E^\beta_- B_{\beta \alpha} ) \approx 0.
\label{4.4}
\end{eqnarray}

%%%%%%%%%%%%%%%%%%%%%%%%%%%%%%%%%%%%%%%%%%%%%%%%%%%%%%%%%%%%%%%%%%%
Under the BRS transformation generated by the BRS charge (\ref{2.11}),
the Green-Schwarz action is transformed as
%**   4.5 %%%%%%%%%%%%%%%%%%%%%%%%%%%%%%%%%%%%%%%%%%%%%%%%%%%%%%%%%
\begin{eqnarray}
\delta I_{GS} =  \int d^2 z \ \lambda \Gamma^a E_{- a} \hat{E}_+,
\label{4.5}
\end{eqnarray}
%%%%%%%%%%%%%%%%%%%%%%%%%%%%%%%%%%%%%%%%%%%%%%%%%%%%%%%%%%%%%%%%%%%
where $\hat{E}^\alpha_+$ is defined  in Eq. (\ref{2.4}).

Following the same procedure as in a flat background, it is easy to
find a new term $I_{new}$ such that a total action $I_0 = I_{GS}
+ I_{new}$ is invariant under the BRS transformation. The new term
takes the form
%**   4.6 %%%%%%%%%%%%%%%%%%%%%%%%%%%%%%%%%%%%%%%%%%%%%%%%%%%%%%%%%
\begin{eqnarray}
I_{new} =  \frac{1}{2} \int d^2 z \ \frac{( d \Gamma^b v^0)
(\lambda \Gamma_b \hat{E}_+)}{v^0 \lambda}.
\label{4.6}
\end{eqnarray}
%%%%%%%%%%%%%%%%%%%%%%%%%%%%%%%%%%%%%%%%%%%%%%%%%%%%%%%%%%%%%%%%%%%
To show that this term transforms as $\delta I_{new} = - \int d^2 z \
(\lambda \Gamma^a \hat{E}_+) E_{- a}$, it is necessary to make use of
$\delta \hat{E}^\alpha_+$ which is given by
%**   4.7 %%%%%%%%%%%%%%%%%%%%%%%%%%%%%%%%%%%%%%%%%%%%%%%%%%%%%%%%%
\begin{eqnarray}
\delta \hat{E}^\alpha_+ &=&  \frac{1}{4} e^\varphi E^b_+ \lambda^\beta
\Big\{ \frac{1}{6} \Big[ ( \Gamma_b \Gamma^{f_1 f_2 f_3} )_\beta{}^\alpha
+  ( \Gamma^{f_1 f_2 f_3} \Gamma_b )_\beta{}^\alpha \Big] T_{f_1 f_2 f_3}
- ( \Gamma^a \Gamma_b )_\beta{}^\alpha D_a \varphi \Big\} \nn\\
&{}& - e^\varphi E^\gamma_+ \lambda^\beta
\Big[ \delta^\sigma_\beta \delta^\alpha_\gamma - \frac{1}{2}
(\Gamma_b)_{\beta\gamma} (\Gamma^b)^{\alpha\sigma} \Big] D_\sigma \varphi
+ e^\varphi \partial_+ \lambda^\alpha.
\label{4.7}
\end{eqnarray}
%%%%%%%%%%%%%%%%%%%%%%%%%%%%%%%%%%%%%%%%%%%%%%%%%%%%%%%%%%%%%%%%%%%
(This equation is also needed to check the nilpotency of the BRS
transformation,
$\delta^2 I_{GS} = 0$.)

Since we have found an invariant action, we can perform the "gauge
fixing" in a standard way. As gauge fermion we choose
%**   4.8 %%%%%%%%%%%%%%%%%%%%%%%%%%%%%%%%%%%%%%%%%%%%%%%%%%%%%%%%%
\begin{eqnarray}
\Psi = \int d^2 z\ \omega_\alpha \hat{E}^\alpha_+.
\label{4.8}
\end{eqnarray}
%%%%%%%%%%%%%%%%%%%%%%%%%%%%%%%%%%%%%%%%%%%%%%%%%%%%%%%%%%%%%%%%%%%
Using Eqs. (\ref{2.17}), (\ref{2.18}), (\ref{4.7})
as well as the identity
%**   4.9 %%%%%%%%%%%%%%%%%%%%%%%%%%%%%%%%%%%%%%%%%%%%%%%%%%%%%%%%%
\begin{eqnarray}
\delta^\sigma_\beta \delta^\alpha_\gamma - \frac{1}{2}
(\Gamma_b)_{\beta\gamma} (\Gamma^b)^{\alpha\sigma}
= - \frac{1}{4} \delta^\alpha_\beta \delta^\sigma_\gamma
- \frac{1}{8} (\Gamma_{f_1 f_2})_\beta{}^\alpha
(\Gamma^{f_1 f_2})_\gamma{}^\sigma,
\label{4.9}
\end{eqnarray}
%%%%%%%%%%%%%%%%%%%%%%%%%%%%%%%%%%%%%%%%%%%%%%%%%%%%%%%%%%%%%%%%%%%
we can evaluate the BRS transformation of the gauge fermion whose
result is given by
%**   4.10 %%%%%%%%%%%%%%%%%%%%%%%%%%%%%%%%%%%%%%%%%%%%%%%%%%%%%%%%%
\begin{eqnarray}
\delta \Psi &=& \int d^2 z \ \Big\{ d_\alpha \Big[ \delta^\alpha_\beta
- \frac{1}{2} \frac{(\Gamma^b \lambda)_\beta ( v^0 \Gamma_b )^\alpha}
{v^0 \lambda}  \Big] \hat{E}^\beta_+ + e^\varphi \Big[ e^{\frac{\varphi}{4}}
\omega \partial_+ (e^{-\frac{\varphi}{4}} \lambda) \Big] \nn\\
&-& \frac{1}{2} e^\varphi
N^{bc} \Big[E^a_+ T_{abc} + E_{+ b} D_c \varphi + \frac{1}{2}
(E_+ \Gamma_{bc} D \varphi) \Big] \Big\}.
\label{4.10}
\end{eqnarray}
%%%%%%%%%%%%%%%%%%%%%%%%%%%%%%%%%%%%%%%%%%%%%%%%%%%%%%%%%%%%%%%%%%%
Then the "gauge-fixed", BRS-invariant action $I = I_{GS} + I_{new} + \delta
\Psi$
takes the form
%**   4.11 %%%%%%%%%%%%%%%%%%%%%%%%%%%%%%%%%%%%%%%%%%%%%%%%%%%%%%%%%
\begin{eqnarray}
I =  I_{GS} + \int d^2 z \  
e^\varphi \Big[ \hat{\omega}_\alpha \partial_+ \hat{\lambda}^\alpha
+ d_\alpha (E^\alpha_+ - \frac{1}{2} E^a_+ \Gamma^{\alpha\beta}_a 
D_\beta \varphi)
+ N_\alpha{}^\beta ( D_\beta \varphi E^\alpha_+
- \frac{1}{2} E^a_+ \hat{T}_{a \beta}{}^\alpha) \Big],
\label{4.11}
\end{eqnarray}
%%%%%%%%%%%%%%%%%%%%%%%%%%%%%%%%%%%%%%%%%%%%%%%%%%%%%%%%%%%%%%%%%%%
where we have defined
%**   4.12 %%%%%%%%%%%%%%%%%%%%%%%%%%%%%%%%%%%%%%%%%%%%%%%%%%%%%%%%%
\begin{eqnarray}
\hat{T}_{a \beta}{}^\alpha \equiv  T_{a \beta}{}^\alpha - \frac{1}{8} 
(\Gamma_a \Gamma^b)_\beta{}^\alpha D_b \varphi,
\label{4.12}
\end{eqnarray}
%%%%%%%%%%%%%%%%%%%%%%%%%%%%%%%%%%%%%%%%%%%%%%%%%%%%%%%%%%%%%%%%%%%
and we have rescaled the antighost $\omega_\alpha$ and the ghost
$\lambda^\alpha$ as
%**   4.13 %%%%%%%%%%%%%%%%%%%%%%%%%%%%%%%%%%%%%%%%%%%%%%%%%%%%%%%%%
\begin{eqnarray}
\hat{\omega}_\alpha = e^{\frac{\varphi}{4}} \omega_\alpha, \
\hat{\lambda}^\alpha = e^{-\frac{\varphi}{4}} \lambda^\alpha.
\label{4.13}
\end{eqnarray}
%%%%%%%%%%%%%%%%%%%%%%%%%%%%%%%%%%%%%%%%%%%%%%%%%%%%%%%%%%%%%%%%%%%
Eq. (\ref{4.11}) is equivalent, modulo superfield redefinitions,   
to a $\sigma$-model action obtained by Berkovits (i.e., Eq. (5.2) 
in Ref. \cite{Ber1}) via a different procedure (and in the case 
of type II superstrings).

%%%%%%%%%%%%%%%%%%%%%%%%%%%%%%%%%%%%%%%%%%%%%%%%%%%%%%%%%%%%%%%%%%%%%
%%%%%%%%%%%%%%%%%%%%%%%%%%%%%%   SEC  5    %%%%%%%%%%%%%%%%%%%%%%%%%%
%%%%%%%%%%%%%%%%%%%%%%%%%%%%%%%%%%%%%%%%%%%%%%%%%%%%%%%%%%%%%%%%%%%%%
\section{Discussions}

In this paper we have presented a reformulation of the Berkovits approach
to the covariant quantization of the GS superstrings, which holds both in flat
and in curved backgrounds. In particular in curved background our
formulation provides a straightforward way to write down the $\sigma$-model 
action.

The method consists of two steps. First one adds to the GS action $I_{GS}$
in conformal gauge a new action term $I_{new}$ to get an action $I_0$ 
invariant under the BRS
transformation generated by $Q_{BRS}$, (\ref{2.11}).
Then one adds to $I_0$ the BRS variation of a suitable gauge fermion, as 
in standard BRS formalism.

$I_{new}$ contains the fields $p_\alpha$ through $d_\alpha$ and the variation
 of $I_0$ with respect to $p_\alpha$ yields the field equation $KE_{+}= 0$
 (i.e. $K\bar{\partial}\theta = 0$ in the flat case). We recall that $K$
is a projector and its trace is given by $tr{K} = 5 $.
Therefore $I_{new}$ can be considered as a sort of partial gauge fixing of
$\kappa$ symmetry which however has the, not obvious, virtue to yield an
action $I_0$
invariant under a BRS symmetry involving a pure spinor of ghosts (eleven
components).

A peculiar feature of this BRS symmetry,
is that it is not related to  a local gauge symmetry as usual 
(in this case with anticommuting parameters). Indeed, anticommuting pure
spinors
do not exist.

 We stress the fact that the invariance under diffeomorphisms of the GS 
 action  has been gauge fixed in conformal gauge    
 $\it{without}$ adding the corresponding $b-c$ ghosts. This is justified by 
 the fact that the central charge vanishes without these ghosts and that the 
 cohomology of the BRS charge ($\ref{2.11}$) is the correct one (see also
 \cite{Ber6}, note 5 in page 9). However in our opinion this point requires 
 a better understanding and deserves further investigation.
 
 A possible problem in our formalism is that the action $I_0 = I_{GS} +
I_{new}$ 
is manifestly $\it{not}$ invariant under the Lorentz transformations.
However, from Eq. (\ref{3.4}) and the fact that $Q_{BRS}$ commutes with the
Lorentz generators, it follows that the Lorentz variation of $I_{new}$ is BRS
invariant. Even more, it is a trivial cocycle of the BRS cohomology.
In fact, the total action $I = I_0 + \delta \Psi$ is Lorentz invariant so
that 
the Lorentz variation of $\delta \Psi$, a trivial cocycle, just compensates
that of
$I_{new}$. 
The fact that the Lorentz variation of $I_0$ is a trivial cocycle
assures us that, in the physical sector,
the theory remains Lorentz invariant despite the non invariance of $I_0$.

It is interesting to notice that, whereas the pure spinor $\lambda$ can be
considered as a covariant object, its conjugate momentum $\omega$ is not so
as a consequence of (\ref{2.17}). However the compound fields $\omega_{\alpha}
\lambda^{\alpha}$ , $\omega_{\alpha} \bar{\partial} \lambda^{\alpha}$ ,
$N^{ab} = \omega \Gamma^{ab} \lambda$ are  covariant tensors unlike
$N^{a_1...a_4} = \omega \Gamma^{a_1...a_4} \lambda$ that does not transform
covariantly.  It is gratifing that the SUGRA constraints prevent the presence
of $N^{a_1..a_4}$ in the final action (\ref{4.11}).

%%%%%%%%%%%%%%%%%%%%%%%%%%%%%%%%%%%%%%%%%%%%%%%%%%%%%%%%%%%%%%%%%%
%%%%%%%%%%%%%%%%%%%%%%%% Acknowledgements %%%%%%%%%%%%%%%%%%%%%%%%%%%%%
%%%%%%%%%%%%%%%%%%%%%%%%%%%%%%%%%%%%%%%%%%%%%%%%%%%%%%%%%%%%%%%%%%
\begin{flushleft}
{\bf Acknowledgements}
\end{flushleft}
We are grateful to I. Bandos, J. Isidro, M. Matone, P. Pasti
and D. Sorokin for stimulating discussions and N. Berkovits for
very useful comments.
Work of the second author (M.T.) has been supported in part by the European
Commission TMR Programme ERBFMPX-CT96-0045. The first author (I.O.) would
like to thank Dipartimento di Fisica, Universita degli Studi di Padova
for its kind hospitality.

\vs 1
%%%%%%%%%%%%%%%%%%%%%%%%%%%%%%%%%%%%%%%%%%%%%%%%%%%%%%%%%%%%%%%%
%%%%%%%%%%%%%%%%%%%%%%%% reference %%%%%%%%%%%%%%%%%%%%%%%%%%%%%%%
%%%%%%%%%%%%%%%%%%%%%%%%%%%%%%%%%%%%%%%%%%%%%%%%%%%%%%%%%%%%%%%%%%

\end{document}